# Copper passivation procedure for water-filled copper cells for applications in metrology

Bruno Buée[*], Alexandre Vergé[*], Vladimir Vidal[*†], Eric Georgin[**], Fernando Sparasci[*]
[*] *LNE-CNAM, La Plaine Saint-Denis, France*
[**] *LNE-CETIAT, Villeurbanne, France*
[†] vladimir.vidal@cnam.fr

## 1. Introduction

In the framework of the European research project MeteoMet [1], the LNE-CNAM [2] and the LNE-CETIAT [3] are developing novel copper cells for the determination of the water vapour pressure equation and the measurement of the triple point of water temperature at the highest degree of accuracy.

The Water Vapour Pressure Equation (WVPE) is the basic formula for the calculation of the vapour pressure and other humidity related quantities. Improvement in the uncertainty of the water vapour formulation in the temperature range between -80 °C (and possibly down to -90 °C) and +100 °C is needed for the improvement of primary standards in the field of hygrometry.

The Triple Point of Water (TPW, temperature 273,16 K) plays a key role in thermometry, because of its threefold function. On one side, it defines the temperature unit, the kelvin. On the other side, it is the reference point in Standard Platinum Resistance Thermometer (SPRT) calibrations, according to the International Temperature Scale of 1990 (ITS-90). Finally, it is the connection point between the current definition of the kelvin and the future one, based on the Boltzmann constant.

Copper cells for WVPE and TPW allow highly accurate temperature control, fast thermal response time and excellent thermal uniformity, because of the elevated copper thermal conductivity. Thus, they can be an outstanding tool for reducing measurement uncertainty. Nevertheless, accurate filling techniques and procedures are necessary, to minimize water contamination and get stable, accurate and reproducible results.

This report describes the filling system and the copper passivation and cleaning procedure, as well as the method for filling copper cells with high-purity water.

## 2. Filling System

In this section, we describe the bench to fill the copper cells. A schematic representation is given in Figure 1

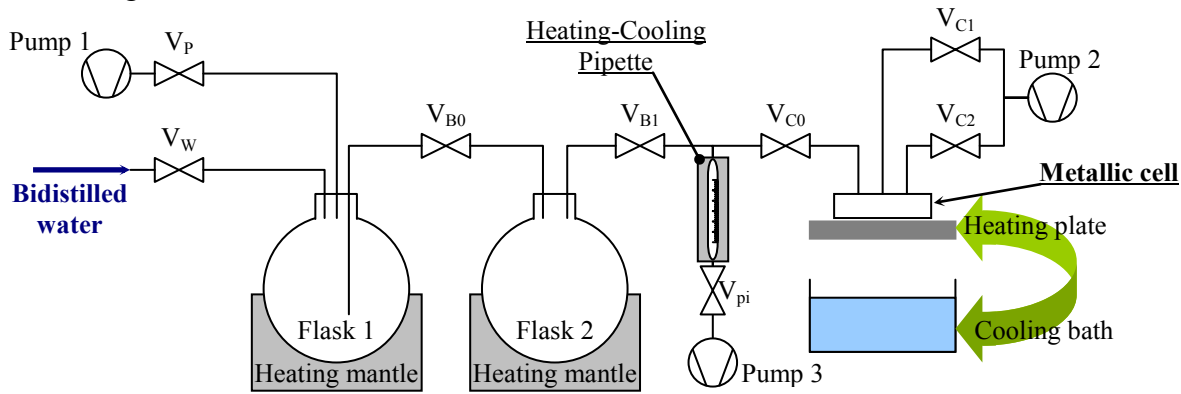

*Figure 1 : Layout of the filling system for copper cells.*

*Bidistiller*
The filling bench consists of a bidistiller, producing a continuous flow of water with sufficient purity, followed by a vacuum-tight, two-stage, outgassing system, which is finally connected to the copper cell. The process is fully realized under vacuum, to suppress any contamination from air.
Water in the bidistiller is supplied by a reverse osmosis plant, providing pure water. The bidistiller operates in air and is kept in continuous operation, so that steam pressure inside minimizes the air flow from the external environment, ensuring the best cleanliness of the parts. In addition, "fresh" bidistilled water is directly supplied to the outgassing system and never stored in any intermediate tank.

*Flask 1*
Bidistilled water flows into a first borosilicate glass flask of 1 l volume, named Flask 1. The flask is provided with a ground-glass stopper made of borosilicate glass, where a PTFE seal ensures the sealing between the ground-glass stopper and the flask [4]. Inlet and outlet glass tubes are welded on the glass caps. Water is admitted from the bidistiller at ambient pressure. When the flask is full, the air inside is pumped and the outgassing process starts. The inlet tube of the Flask 1 is flush to the inner surface of the cap and the outlet tube dips in the bottom of the flask. A heating mantle, providing 350 W over 1 l volume, is used to heat the water up to the boiling temperature.

*Flask 2*
After outgassing, water flows through the outlet tube of Flask 1 to reach a second flask, named Flask 2, identical to the previous, except for the outlet tube which is flush to the inner surface of the glass stopper. Flask 2 is never exposed to air, and the transfer from Flask 1 to Flask 2 is performed either under vacuum, or under water steam atmosphere. Temperature of Flask 2 can be raised with a heating mantle identical to that mounted on Flask 1.

---
[4] Caps will be glass-welded to the flasks in future versions of the system and PTFE seals will be removed.

*Copper cell*
The copper water cell has two or more outlet tubes [5] and one inlet connected to the bench. The cell may be either placed on a heating plate whose temperature can range from 3 °C above room temperature to 300 °C with ± 2% stability, or in bath at about 0 °C, according to the procedure in progress (either cleaning or filling).

*Other elements*
A graduated pipette (in borosilicate glass) is connected to the input of the cell to estimate the amount of liquid water in the cell during the filling. As for the cell, the pipette can be either heated or cooled.
Three primary vacuum pumps (shown in Figure 1) are used to induce a vacuum of about 20 mbar at 20 °C.
The tubes connecting the elements of the bench are made of glass, with 5 mm internal diameter and 8 mm outer diameter.
All the elements represented as valves in Figure 1 are silicone tubes of 4 mm internal diameter and 10 mm outer diameter, with lengths ranging from few centimetres to few tens of centimetres. They can be clamped to act as closed valves and have the advantage of minimizing the pollution induced by classical metal or glass valves.

## 3. Copper cell charactheristics

Vapour pressure and triple point cells are constructed with copper. Since copper is not soluble in water, the water pollution by copper can be excluded. However the cuprous ions are slightly [pal04] soluble and the presence of soluble oxygen in solid copper [hor77][nar83] promotes the formation of cuprous ions [dor01]. Therefore it is necessary to reduce the concentration of oxygen in copper.
It is recommended to use Oxygen Free High Conductivity (OFHC) copper, because it is manufactured under controlled and neutral atmosphere, which strongly reduces the presence of oxygen and improves thermal conductivity [anc]. In the same goal, cell sealing is performed by electron beam welding [dav93], in order to not bring any additional impurity in the OFHC copper.

## 4. Filling procedure

The filling procedure consists of four major phases, to ensure the purity of the water in the cells: bench cleaning, cell cleaning, cell filling. They are subdivided in several steps and presented in the next sections. Figures 2 to 5 provide an overview of the steps in the filling procedure.

### Phase 1: Bench cleaning

*Refer to Figure 2.*

During bench cleaning, the cell is not yet connected to the bench.

- A. The cell is disconnected. Air in the filling bench is evacuated with primary pumps (1 & 2), to reduce the concentration of water soluble gas.

---

[5] Cells are designed with redundant outlets to preserve the possibility of carrying out additional experiments, like water doping with controlled amounts of impurities. At the present time the outlets are connected to the vacuum pump and are sealed at the same time after the filling.

B. The bench is rinsed with bidistilled water to remove possible water soluble elements. It is necessary to employ at least 9 l of bidistilled water to get a good rinsing. At the end, the flasks are kept full of bidistilled water.
C. $N_2$, $O_2$, $CO_2$ and other gases are degassed from water in Flask 1 and Flask 2. Water in the flasks is heated close to the boiling temperature of water and degassed with a primary pump. This has the effect of removing the dissolved gases in water and replaces air in the bench by water vapour. The pumping is only primary, but the pump flow associated with the steam flow ensures total evacuation of gaseous impurities in the bench.
D. In previous step, the heating-degassing of the water can lower the water level in the flasks. It is then necessary to refill Flask 2 with pure water without introducing impurities. Flask 2 is filled with outgassed water from Flask 1. The process is performed under vacuum, without introducing air in the entire bench.
Valves $V_P$ and $V_W$ are closed. Pumping is carried out on $V_{pi}$ with opened valves $V_{B0}$ and $V_{B1}$, while Flask 1 is heated close to the boiling point. The heating on Flask 1 and the pumping push the water from Flask 1 to Flask 2.

## Phase 2: Flasks filling

*Refer to Figure 3.*

E. Although the amount of water required to fill the cell is low, cleaning and outgassing consume water. It is therefore necessary to fill the flasks without introducing impurities into the bench. It is then necessary to degas the water newly introduced.
   a. Filling of Flask 1: valve $V_{B0}$ is closed to not introduce not-degassed water in the bench. Flask 1 is filled with bidistilled water via $V_W$. The pump n. 1 is activated on the opened valve $V_P$ to facilitate the filling and to avoid introducing air and impurities.
   b. Degassing of Flask 1: valve $V_{B0}$ is still closed, the pump n. 1 is still running and $V_P$ is open. The valve $V_W$ is closed and Flask 1 is heated until water level decreases by one third.
   c. Filling of Flask 2: This step is equivalent to the step D. With valves $V_P$ and $V_{C0}$ closed and valves $V_{pi}$, $V_{B0}$ and $V_{B1}$ opened, Flask 1 is heated to increase the pressure and push the degassed water to Flask 2.

Step E can optionally be repeated more times, as soon as the flasks are empty.

## Phase 3: Cell cleaning

*Refer to Figure 4.*

F. Copper cell degassing:
To degas the copper constituting the cell, the cell is heated for more than six hours under vacuum at about 60 °C, by means of the heating plate. This step can be performed at the same time as steps C and D.
G. Prior to start this step, Flask 1 and Flask 2 must be full of bidistilled and degassed water. Flask 1 is heated to increase its inner pressure and push the water of the full Flask 2 inside the bench and the cell. This step is performed to rinse the cell with bidistilled and degassed water, in order to dissolve and remove any soluble element that could be still present in the cell.

H. Last stage before cell filling with bidistilled water: cell draining. The steam flow associated with pumping removes all the water present in the cell. The concomitant heating of the cell leaves only steam inside.

Steps F, G and H are repeated at least 2 times before starting the cell filling phase.

**Phase 4: Cell filling**
*Refer to Figure 5.*

I. The filling of the cell is performed by distillation.
    a. The graduated pipette is cooled and so filled by condensation of the evaporated water from Flask 2, kept under heating.
    b. Then the water in the pipette is evaporated and condensed in the cooled cell. The graduated pipette allows knowing the amount of water that is condensed in the cell. When the desired amount of water is evaporated, inlet and outlet tubes are permanently sealed by pinching-off [her03].

## 5. Conclusion

The bench realized by the LNE-CNAM allows filling copper cells for WVPE and TPW measurements with high-purity water.
The validation of the system and the procedure is in progress: measurements on recently-filled WVPE and TPW cell prototypes are being realized, in order to check the presence of impurities. Results will be published in next papers. They will be useful either to assess the validity of the procedure presented here, or to propose amendments in future versions.

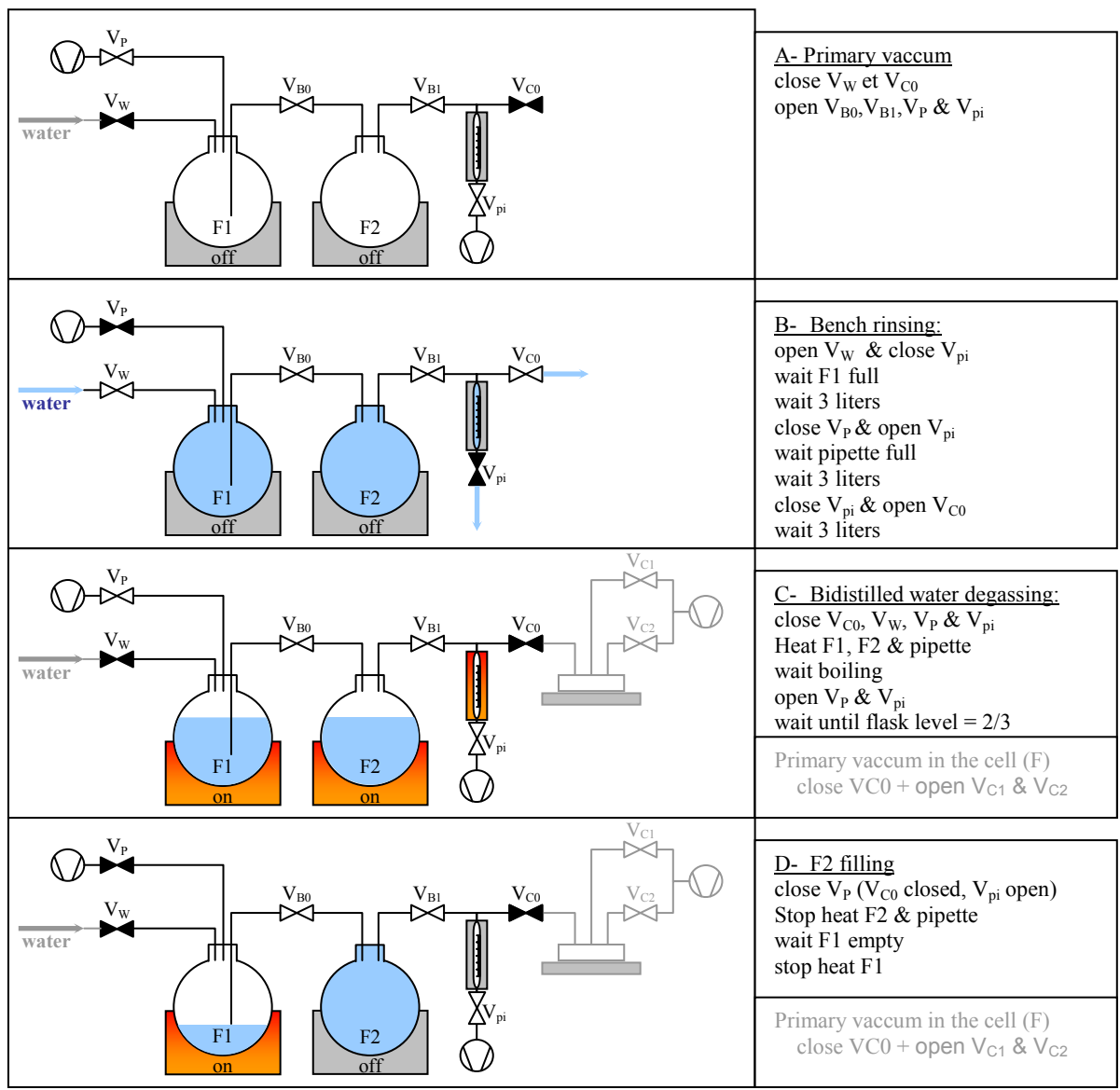

*Figure 2: **Bench cleaning protocol**. Solid black valves are closed valves, white valves are open valves, red/orange areas represent powered heaters.*

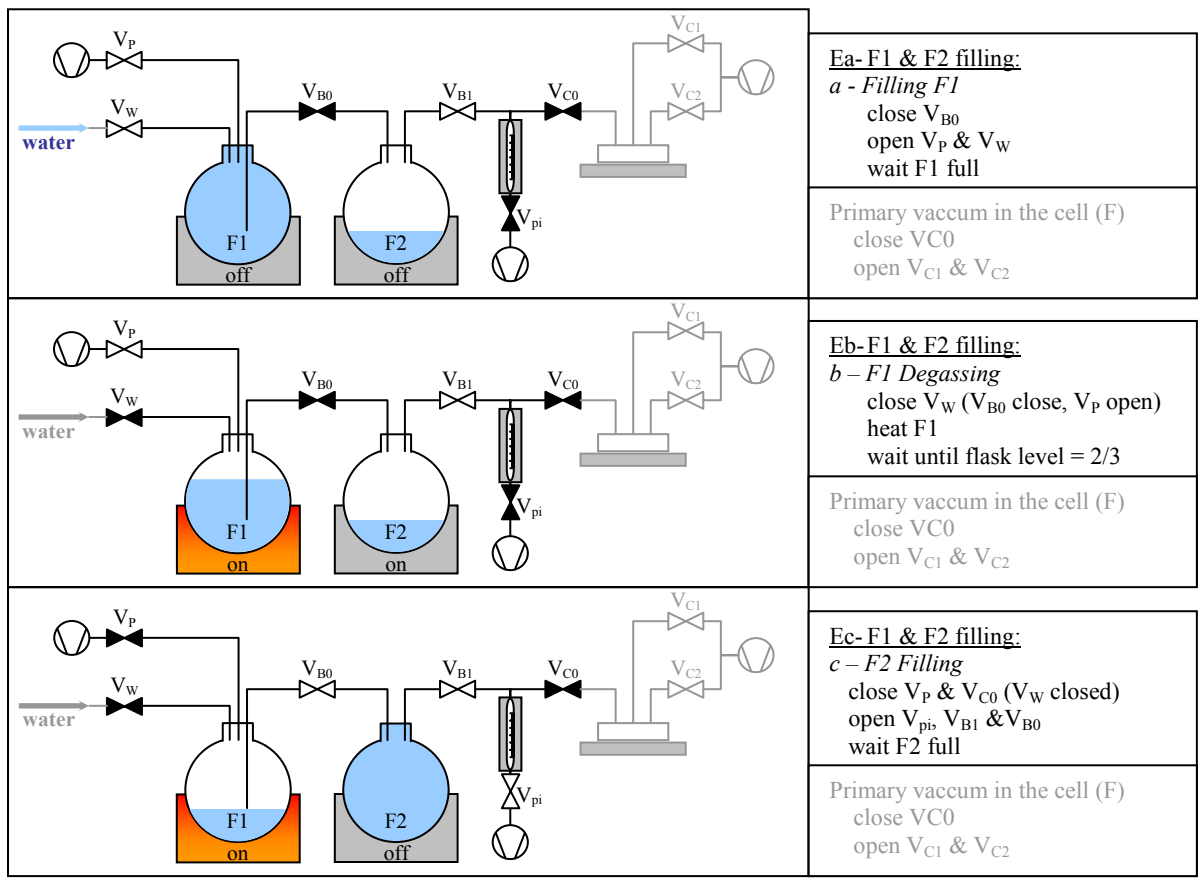

*Figure 3: **Flasks filling protocol**. Solid black valves are closed valves, white valves are open valves, red/orange areas represent powered heaters.*

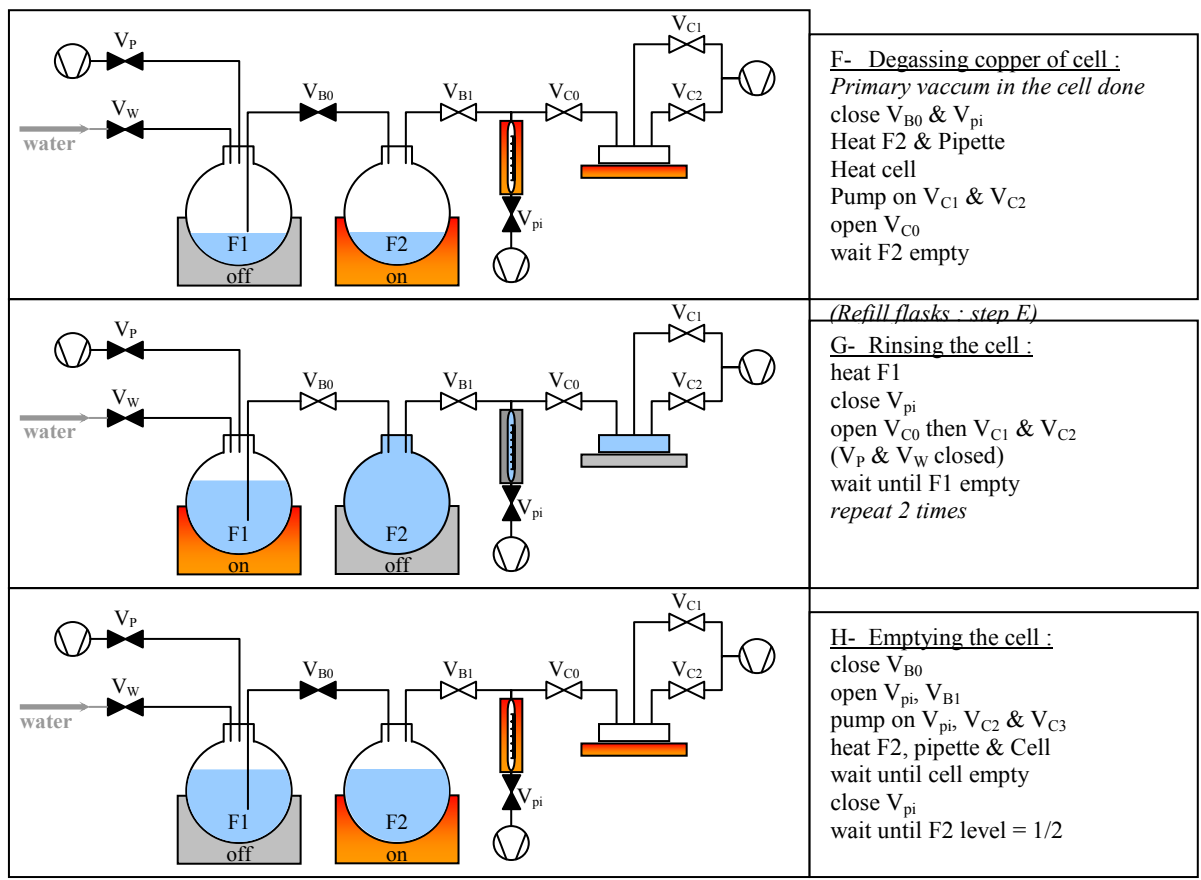

*Figure 4:* **Cell Cleaning protocol**. *Solid black valves are closed valves, white valves are open valves, red/orange areas represent powered heaters.*

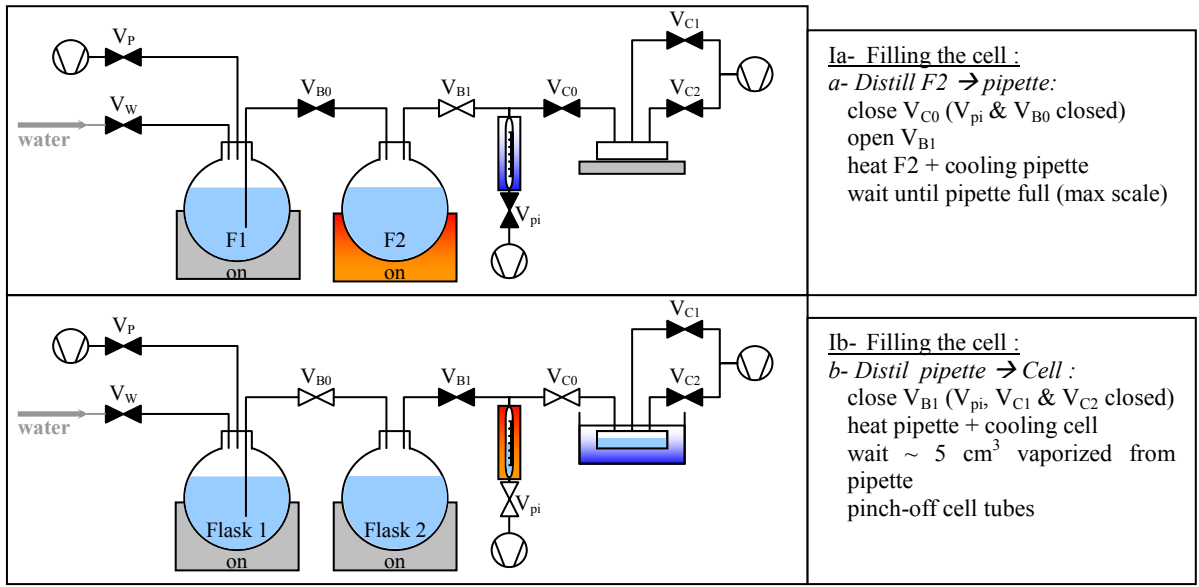

*Figure 5: **Cell filling protocol**. Solid black valves are closed valves, white valves are open valves, red/orange areas represent powered heaters.*